\documentclass[1p]{elsarticle}
\usepackage{xcolor}
\usepackage{lineno,hyperref}
\usepackage{amsmath}
\usepackage{mathtools}
\usepackage[]{tikz}

\DeclareMathOperator{\Tr}{Tr}

\begin{document}
\begin{frontmatter}
%Title of paper
\title{Wilson loop of the heterotic sigma model and the sv-map}

\author[new,uw]{Wei Fan}
\address[neu] {Department of Physics,	Northeastern University, Boston, MA 02115, USA}
\address[uw]{ Institute of Theoretical Physics, Faculty of Physics,	University of Warsaw, Poland}

\date{\today}

\begin{abstract}
The  single-valued projection (sv) is  a relation between scattering amplitudes of gauge bosons in heterotic and open superstring theories. Recently we have studied sv from the aspect of nonlinear sigma models~\cite{FFST2018195}, where the gauge physics of  open string sigma model is under the Wilson loop representation but the gauge physics of  heterotic string sigma model is under the fermionic representation since the Wilson loop representation is absent in the heterotic case. There we showed that the sv comes from a sum of six radial orderings of heterotic vertices on the complex plane. In this paper, we propose a Wilson loop representation for the heterotic case and using the Wilson loop representation to show that sv comes from a sum of two opposite-directed contours of the heterotic sigma model. We firstly prove that the Wilson loop is the exact propagator of the fermion field that carry the gauge physics of the heterotic string in the fermionic representation. Then we construct  the action of the heterotic string sigma model in terms of the Wilson loop, by  exploring the geometry of the Wilson loop and by generalizing the nonabelian Stokes's theorem~\cite{nonabelianStokes1,nonabelianStokes2,nonabelianStokes3} to the fermionic case.  After that, we compute some three loop and four loop diagrams as an example, to show how the sv  for  $\zeta_2$ and $\zeta_3$  arises from a sum of the contours of the Wilson loop. Finally we conjecture that this sum of contours of the Wilson loop is the mechanism behind the sv for general cases. 
\end{abstract}


\end{frontmatter}

\section{Introduction}

For tree-level string amplitudes, the single-valued projection (sv)~\cite{Brown:2013gia} is a map  between gluon amplitudes of the open superstring and gluon
amplitudes of  heterotic string~\cite{StiebergerSVmath14,StiebergerTaylorSVP,Stieberger:2016xhs}. For some recent proof see~\cite{brownNew18,Schlotterer:2018zce,Stieberger:2018edy}. These tree-level string amplitudes are expressed in terms of multiple Gaussian  hypergeometric functions, which contains the parameter of $\alpha'$~\cite{stiebergerEulerZagier}. The Taylor expansion of them in terms of small $\alpha'$ contains coefficients of multiple zeta values (MZV) at each order. Focusing on the single-trace part of the gluon amplitudes, when we take the expansion for the open superstring case and perform the sv on the MZV of the coefficients, the result is directly the expansion of amplitudes of the heterotic string case. 

Generically, in the $\alpha'$-expansion of open superstring tree-level amplitudes the whole
space of MZVs enters~\cite{PhysRevLett.106.111601,Schlotterer:2012ny}, while closed superstring tree-level amplitudes exhibit
only the subset of SVMZVs in their $\alpha'$-expansion~\cite{PhysRevLett.106.111601,StiebergerSVmath14}. The relation between
open and closed string amplitudes through the the single-valued projection has been
observed in~\cite{StiebergerSVmath14} and established in~\cite{StiebergerTaylorSVP}.

Since all the information are encoded in the hypergeometric functions once for all, it's hard to
see the detailed origin or mechanism behind this sv. We need to go to the nonlinear sigma model approach to investigate its origin for each MZV at each order of $\alpha'$ expansion. We need to compute Feynman diagrams corresponding to single-trace gauge terms on the world-sheet, where the loop number of diagrams of the sigma model corresponds to the order of $\alpha'$ expansion of string amplitudes.  We have studied sv from this aspect in the previous paper~\cite{FFST2018195}, and proposed a sv-map which states that the sv works on the Feynman diagram level for the corresponding sigma models. 

The gauge physics of the open string sigma model can be studied in terms of the 
Wilson loop representation~\cite{Dorn:1986dt, FradkinTseytlin}. The Wilson loop
 is directly gauge covariant, but is hard for the perturbation
calculation because the path ordering nature of the Wilson loop is highly
nontrivial~\cite{WilsonLoopExpansion,BrecherPerry}.  Since the gauge physics comes from the Chan-Paton factor on the boundary of the open string world-sheet, it is a purely 1D problem  and the Wilson loop can be rewritten as a functional integral of a pair of auxiliary Grassmann fields~\cite{Wilsonloop1DFermionic,PhysRevD.19.1153}.  The Wilson line between two points is just the exact propagator of the Grassmann fields evaluated at those two points. The perturbation calculation is straightforward in this fermionic representation, but gauge covariance becomes nontrivial and hard to deal with.

The gauge physics of the heterotic string sigma model lives on the whole 2D world-sheet, and currently can only be described by the fermionic representation. The Wilson loop representation is still missing.  Although the perturbation calculation is
straightforward in the fermionic representation~\cite{Sen:1985eb}, the result of each Feynman diagram is not
gauge covariant. The gauge covariance is hard to deal with, because by definition the model only has superconformal symmetry and does not have the gauge symmetry. A gauge covariant perturbation process was given in~\cite{Ellwanger:1988cc}, but it involves a very specific nonlocal field redefinition procedure.  In~\cite{FFST2018195},
we proposed a method of reorganized perturbation for the fermionic representation to put the perturbation in a gauge covariant manner. 

\begin{figure}
  \centering
  \includegraphics[width=14cm]{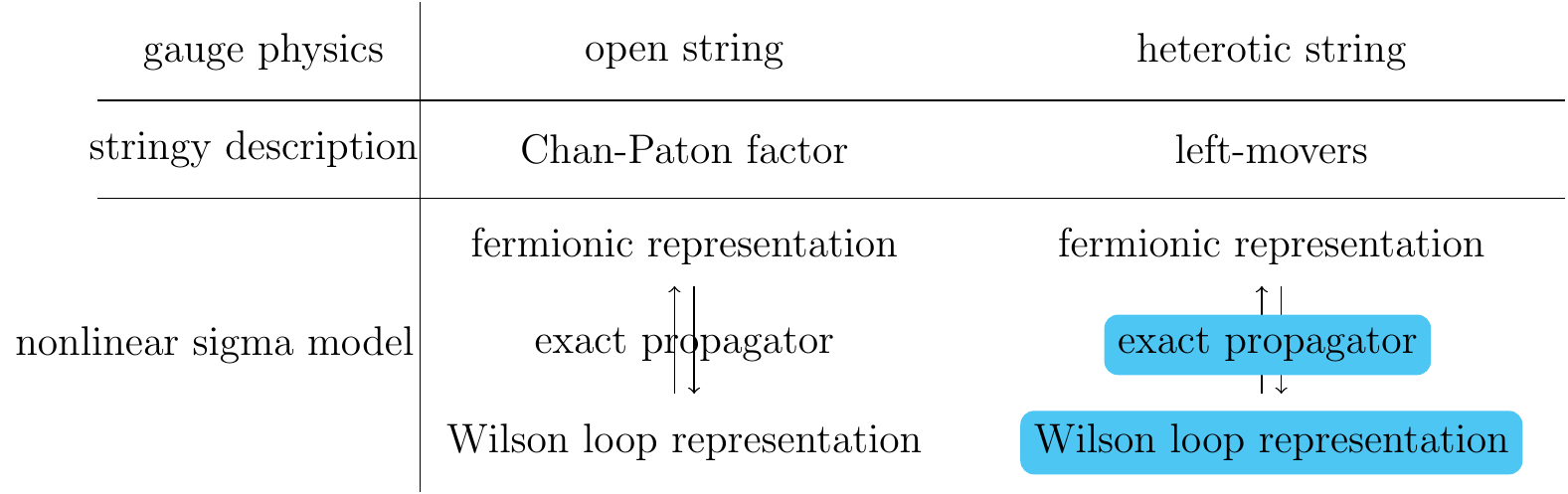}
  \caption{The correspondence of descriptions of gauge physics between the open string and the
    heterotic strings. In string theory, the gauge physics in the open string is carried by the Chan-Paton factors and the gauge physics of the heterotic string is carried by the left-movers. In nonlinear sigma model approach, the fermionic representation and the Wilson loop representation are connected by the fact that the Wilson line is the exact propagator of the fermionic field, both for the 1D boundary of the world-sheet of the open string case and for the 2D world-sheet of the heterotic string case. {\color{cyan}The blue part is the Wilson loop representation proposed in this paper}.}
  \label{correspondence_openHete}
\end{figure}

However,  the absence of a Wilson loop representation for the heterotic string sigma model is still a very bothering fact, because the open string sigma model has it. So in this paper, we will propose a Wilson loop representation
for the gauge physics of the heterotic sigma model. It turns out that the Wilson line between two points on the 2D world-sheet is also the exact propagator of the fermionic field, just like the case of the open string. So now we have a complete correspondence of descriptions of  gauge physics between the open string and the heterotic string, as shown in figure~\ref{correspondence_openHete}. 

Furthermore, after the gauge physics of both the open and the heterotic string sigma model are put under the Wilson loop representation, the sv-map~\cite{FFST2018195} between them turns out to have a very simple geometric origin: the sv-map comes from the sum of two path-ordered integrals of opposite directions. The Wilson loop is path-ordered, so the Feynman integral of gauge physics is a path-order integral. For the open string case, the integration contour is just the boundary itself and we can compute the Feynman integral along this contour. For the heterotic case, we will show that the integration contour is a sum of two opposite-directed contours. When we sum the Feynman integral along these two opposite-directed contours, we will obtain a result which is the sv-map of the corresponding result of the open string case. In the previous paper~\cite{FFST2018195}, we showed that the sv-map comes from a sum of six radial orderings of heterotic vertices on the complex plane when the gauge physics of the heterotic sigma model is not under the Wilson loop representation. In this paper, we show that the sv-map comes from a sum of two opposite-directed contours, using the Wilson loop representation. So the geometric origin is simpler. 
 To show this perturbative calculation,  we firstly need to compute the functional derivative of the Wilson loop for the heterotic string case, which is just an analog of the open string case~\cite{WilsonLoopExpansion}. Then we get  the background field expansion of the Wilson loop for the heterotic string case, which corresponds consistently (in the sense of exact propagators) to the background field expansion of the fermionic representation obtained using reorganized perturbation in our previous paper~\cite{FFST2018195}. Finally we will compute several diagrams as an example to show this geometric picture.  

This paper is organized as following. In section~\ref{sec:openString}, we will briefly
review the open string sigma model in the fermionic representation and in the
Wilson loop representation. In
section~\ref{sec:heteString}, we will study the heterotic string sigma model. Firstly, we will briefly review
the reorganized perturbation method in our previous paper~\cite{FFST2018195} for the fermionic representation. Then we will construct the Wilson loop, compute its functional derivatives and prove that it is the exact propagator of the fermion field of the fermonic representation. After that, we will investigate the geometry of the Wilson loop and generalize the nonabelian Stokes's theorem to the fermionic case, from which we construct the heterotic sigma model action using the Wilson loop. In
section~\ref{sec:singleValuedMap}, we will explore the connection between the
geometry of the Wilson loop and the sv-map. We will compute several diagrams of three loop and four loop as an example and  give the conjecture for the general case. 

\section{Open superstring}
 \label{sec:openString}

In this section, we briefly review the fermionic representation and
the Wilson loop representation of the gauge physics of the open string sigma
model. This is just for the purpose of completeness, 
to be compared with the heterotic sigma model in section~\ref{sec:heteString}. We are
not going to talk about any perturbation calculations here. 

\subsection{Wilson loop representation}
\label{sec:openWilsonLoop}

For the open string, the gauge degrees of freedom is manifested via
the Chan-Paton factor and only lives on the boundary $\partial\Sigma$ of the open string world-sheet $\Sigma$, which is just
a 1-dim curve (either the unit circle or the real axis depending on the choice)
parameterized by $\tau$. The Wilson loop is directly parametrized by this 1-dim
curve itself. The sigma model  action can be written in a covariant manner on the Euclidean world-sheet as 
\begin{align}
  \label{eq:openWilsonLoop}
  S &= S_\Sigma + S_{\partial\Sigma} \nonumber\\
  S_\Sigma &= \frac{1}{4 \pi \alpha'} \int d^2\sigma_E (\partial X^\mu
               \bar{\partial} X_\mu + \Phi^\mu \bar{\partial} \Phi_\mu +
               \tilde{\Phi}^\mu \partial \tilde{\Phi}_\mu ) \nonumber\\
  S_{\partial\Sigma} &= \ln \Tr \mathcal{P} \exp \{i  \oint_{\partial\Sigma}
                         d\tau (A_\mu (X)\partial_\tau X^\mu - \frac{1}{2}
                         \phi^\mu \phi^\nu F_{\mu\nu})\} \nonumber\\
    &= \ln \Tr \mathcal{P}W[A],  
\end{align}
where $\phi = \Phi\vert_\Sigma = \tilde{\Phi}\vert_\Sigma$ is the fermionic string on the boundary and the open superstring part $S_\Sigma$ follows from~\cite[Chapter~12.3]{polchinsk:string} and the Chan-Paton part $S_{\partial\Sigma}$ is defined
in~\cite{Tseytlin:1997csa}.  

The Wilson loop is  $W[A] = \Tr V[X,\phi,\tau, \tau]$ where
$V[X,\phi,\tau_2,\tau_1]$ is the Wilson line defined as
\begin{equation}
\label{eq:openWilsonLine}
  V[X,\phi,\tau_2,\tau_1] \coloneqq \mathcal{P} e^{i  \int_{\tau_1}^{\tau_2} d\tau [A_\mu (X)\partial_\tau X^\mu - \frac{1}{2}
\phi^\mu \phi^\nu F_{\mu\nu}]}
\end{equation}
For convenience, we will omit the path ordering symbol $\mathcal{P}$ from now
on. Whenever we deal with the Wilson loop, the path ordering is always implicitly
there.

The functional variation~\cite{WilsonLoopExpansion} of the bosonic part of the Wilson loop under $X\to X + \delta X$ is
\begin{align}
\label{eq:openVariationWilsonBosonic}
V[X+\delta X, \tau_0, \tau_0] &=  e^{i  \oint d\tau [A_\mu (X+\delta X)\partial_\tau (X+\delta X)^\mu} \nonumber \\
&= V[X] - i \delta X^\mu(\tau_0) A_\mu(X(\tau_0)) V[X] +i V[X] A_\mu(X(\tau_0)) \delta X^\mu(\tau_0)  \nonumber \\
&+i \oint d\tau V[X, \tau_0, \tau] F_{\mu\nu} \partial_\tau X^\nu(\tau) \delta X^\mu(\tau) V[X, \tau, \tau_0].
\end{align}
 Combined with the fermionic part it gives~\cite{BrecherPerry} the background field expansion of the action under $X \to X  + \xi$
\begin{align}
  \label{eq:openBGEWilson}
   S_{\partial\Sigma}[X+\xi,\phi] &=  i \oint d\tau \Tr V[X] \{ (A_\mu (X)\partial_\tau X^\mu  + \partial X^\nu [F_{
                    \mu_1\nu} \xi^{\mu_1}+ \sum_{n=2}\frac{1}{n!} D_{\mu_n}\ldots
       D_{\mu_2} F_{ \mu_1\nu} \xi^{\mu_1}\ldots \xi^{\mu_n} ] \nonumber \\
  &{}+[ \frac{1}{2} F_{ \mu_1\mu_2} \xi^{\mu_1} \partial \xi^{\mu_2} +\sum_{n=3} \frac{n-1}{n!} D_{\mu_{n-1}}\ldots
       D_{\mu_2} F_{ \mu_1\mu_n} \xi^{\mu_1}\ldots \xi^{\mu_{n-1}} \partial
       \xi^{\mu_n}] \nonumber\\
  &{} - \frac{1}{2} [F_{\nu_1 \nu_2} \phi^{\nu_1}  \phi^{\nu_2} +
       \sum_{n=1}\frac{1}{n!}  D_{\mu_n}\ldots D_{\mu_1}F_{\nu_1 \nu_2}(X)
       \phi^{\nu_1}\phi^{\nu_2}  \xi^{\mu_1}\ldots \xi^{\mu_n} ] \},
\end{align}
where $\xi$ and $\phi$ are treated as quantum fields. 
\subsection{Fermionic representation}
\label{sec:openFermionic}

The path ordering nature of the Wilson loop can be written in terms of the Heaviside step function. Since this is a 1D problem, the Heaviside step function can be viewed as the free propagator of a pair of fermionic coordinates~\cite{Dorn:1986dt,Arefeva:1980zd}. Then the Wilson loop of the open string  can be viewed as coming from the following ordinary action
\begin{align}
  \label{eq:openFermionic}
  S_{\partial\Sigma}=\oint d\tau \bar{\psi}(\tau) \left(\frac{d}{d\tau} - i (A_\mu (X)\partial_\tau X^\mu - \frac{1}{2}
\phi^\mu \phi^\nu F_{\mu\nu}) \right) \psi(\tau) 
\end{align}
where the pair of fermion coordinates $\psi(\tau), \bar{\psi}(\tau)$ live on the boundary\cite{Dorn:1986dt}.
The Wilson line $V[A]$  is just the exact propagator of this fermion coordinate~\cite{Wilsonloop1DFermionic}
\begin{equation}
  \label{eq:openPhasePropagator}
    V[X,\phi,\tau_2,\tau_1] = \langle \psi(\tau_2) \bar{\psi}(\tau_1) \rangle. 
\end{equation}

\section{Heterotic string}
\label{sec:heteString}

In this section, we will study the gauge physics of the heterotic string sigma model. 
For the heterotic string~\cite{Gross:1985fr}, the gauge physics are generated by  the $16$ left-movers. By analog of the bosonization of two fermion fields, the gauge physics can be described by $32$ real, anticommuting, left-moving, right-handed coordinates  $\psi^j$ which transform under the fundamental representation of $SO(32)$. These coordinates are  Majorana-Weyl spinors on the world-sheet and we will just call them fermion fields for simplicity. This is the fermionic representation  of the heterotic sigma model and the action  is given as~\cite[Chapter~12.~3]{polchinsk:string} 
\begin{align}
  \label{eq:heteFermionicEuclideanSen}
   \mathcal{S}_E = \frac{1}{2 \pi \alpha'} \int d^2z \{  \partial X^{\mu }\bar{\partial }X_{\mu } 
+\phi ^{\mu } \bar{\partial }\phi _{\mu }+\psi^j\partial \psi ^j -i \psi \left(\partial X^{\nu }A_{\nu } - \frac{1}{2}F_{\nu _1v_2}\phi ^{\nu_1}\phi ^{\nu _2}\right)\psi  \},
\end{align}
where the equal time contour on the complex plane is the circle and $\phi^\mu$ is the super partner of $X^\mu$.
This corresponds to the fermionic representation of the open string sigma model eq.~\eqref{eq:openFermionic}, except that the gauge terms here is a 2D integral while in the open string case it is a 1D integral .

\subsection{Reorganized perturbation method}
\label{sec:heteFermionic}

The advantage of the fermionic representation is that the perturbation
calculation is straightforward. However, this perturbation process is not
gauge invariant for each diagram, because of the presence of the term $\partial A_\mu$ in the
action. Only after combining all the diagrams at each loop can we  get a gauge invariant result. If one is doing the complete renormalization, then this does not affect the final result. But if one just want to compute a subset of all the diagrams at each loop, then the lack of gauge invariant would be a trouble. In this case, a perturbation method that is gauge invariant at each diagram level is needed. In~\cite{Ellwanger:1988cc}, a specially chosen nonlocal field transformation is used to put the perturbation into a gauge invariant form at each term. In~\cite{FFST2018195}, we use a reorganized perturbation method to achieve a gauge invariant perturbation without going through the nonlocal field transformation. We briefly recall this reorganized perturbation here, for the purpose of completeness and to be compared with the result of Wilson loop construction in  next section~\ref{sec:heteWilson}.

Usually in the  perturbation calculation
\begin{equation}
  \label{eq:ordinaryPerturb}
  e^{-S(F_{\mu\nu})} = \int DX D\phi  D\psi  e^{-S_E[X,A,\phi, \psi]}, 
\end{equation}
we compute diagrams involving all the propagators of $X, \phi, \psi$ and combine all the
diagrams at each loop to get a gauge invariant result. On the other hand, if we firstly  
integrate out the fermion fields $\psi$ 
\begin{align}
  \label{eq:effectiveActionCalculation}
 e^{-S_{eff}(X, F, \phi)} = \int D\psi  e^{-S_E[X,A,\phi,\psi]},
\end{align}
we would  get a gauge invariant effective action of the gauge field strength
 $F_{\mu\nu}$. Then we can do the remaining functional integral perturbatively in a gauge invariant manner. But in practice, it is impossible to obtain a closed form for $S_{eff}$ perturbatively, since we can not integrating out the $\psi$ field in this brute force manner. Actually this effective action $S_{eff}(X, F, \phi)$ is exactly the Wilson loop representation we want to construct, which corresponds to the open string case eq.~\eqref{eq:openWilsonLoop}.  In the next section~\ref{sec:heteWilson}, we will use an indirect way to get the Wilson loop representation, and show that eq.~\eqref{eq:heteConnectionWilFer} the Wilson line is the exact propagator of the fermion field, like the open string case eq.~\eqref{eq:openPhasePropagator}.  

Our reorganized perturbation method is as following: firstly, we do the
background field expansion $X \to X + \xi$ and treat $\xi, \phi,\psi$ as the quantum field; then we pick all the tree-level diagrams at each order of $\alpha'$; finally we integrate out all the internal $\psi$ propagators and get a gauge invariant results $S(X + \xi, \phi, \psi, F_{\mu\nu})$. 
Now the perturbation calculation is gauge invariant at each diagram.

The background field expansion of the gauge physics using the reorganized
perturbation method~\cite{FFST2018195} is as follows
\begin{align}
  \label{eq:heteReorPer}
   \mathcal{S}_E &= \frac{2}{4 \pi \alpha'} \int d^2\sigma_E\{  \partial X^{\mu
  }\bar{\partial }X_{\mu } + \phi ^{\mu } \bar{\partial }\phi _{\mu
  } + \psi^j\partial \psi ^j   \nonumber \\
    & - i \psi \left( \partial X^{\nu }A_{\nu } +  \partial X^\nu [F_{
                    \mu_1\nu} \xi^{\mu_1}+ \sum_{n=2}\frac{1}{n!} D_{\mu_n}\ldots
       D_{\mu_2} F_{ \mu_1\nu} \xi^{\mu_1}\ldots \xi^{\mu_n} ]\right. \nonumber \\
  &{} \left.+[ \frac{1}{2} F_{ \mu_1\mu_2} \xi^{\mu_1} \partial \xi^{\mu_2} +\sum_{n=3} \frac{n-1}{n!} D_{\mu_{n-1}}\ldots
       D_{\mu_2} F_{ \mu_1\mu_n} \xi^{\mu_1}\ldots \xi^{\mu_{n-1}} \partial
       \xi^{\mu_n}] \right. \nonumber\\
  &{} \left.-\frac{1}{2} [F_{\nu_1 \nu_2} \phi^{\nu_1}  \phi^{\nu_2} +
       \sum_{n=1}\frac{1}{n!}  D_{\mu_n}\ldots D_{\mu_1}F_{\nu_1 \nu_2}(X)
       \phi^{\nu_1}\phi^{\nu_2}  \xi^{\mu_1}\ldots \xi^{\mu_n} ] \right)\psi\}.
\end{align}
The gauge terms here have the same structure as background field expansion of the Wilson loop in the
open string case eq.~\eqref{eq:openBGEWilson}, except that for the heterotic string
the gauge terms are carried by the fermion fields $\psi$.

\subsection{Construct the Wilson loop}
\label{sec:heteWilson}

In this section, we will build up the Wilson loop  of the gauge terms of 
the heterotic string sigma model, and show that  it satisfies all the
requirements. However, this is just a single quantity of Wilson loop, not the complete action.  (In the next section~\ref{sec:geoWilson}, we will propose a way to rewrite the action of the heterotic sigma model using this Wilson loop, based on analyzing the geometry of the Wilson loop.)

So in this section we
will only consider gauge terms of the heterotic sigma model
\begin{align}
  \label{eq:heteFermPart}
  L_E^f[A, \psi] &=  \psi D_z \psi + \frac{i}{2} \psi F_{\nu_1 \nu_2}
                     \phi^{\nu_1} \phi^{\nu_2} \psi \nonumber \nonumber \\ 
  &=  \psi \partial_z \psi -i \psi (A_\mu \partial_z X^\mu - \frac{1}{2}
      F_{\nu_1 \nu_2} \phi^{\nu_1} \phi^{\nu_2} ) \psi.  
\end{align}
The notation for the gauge fields are 
$A_\mu = A^a_\mu T^a$, $ D_\mu = \partial_\mu - i \left[ A_\mu, \cdot \right]$ and 
 $D_z = \partial_z - i \left[ A_z,  \cdot \right]$.  
Classically this Lagrangian has the gauge symmetry under the following transformation
\begin{align}
  \label{eq:heteGaugeTransf}
  A_\mu &\rightarrow U A_\mu U^\dagger + i U \partial_\mu U^\dagger \nonumber
  \\
  \psi &\rightarrow U \psi,
\end{align}
which leads to $F_{\mu \nu} \rightarrow U F_{\mu \nu} U^\dagger$, 
  $D_z\psi \rightarrow U D_z\psi$ and 
  $L_E^f[A,\psi] \rightarrow L_E^f[A, \psi]$.  
Notice that the heterotic string sigma model is only required to have superconformal symmetry. The gauge symmetry defined above is just a field redefinition from the point of view of the world-sheet. That's why we did not have gauge invariant results for each diagram in the perturbation calculation. Only after integrating out all the world-sheet, we get the space-time effective action of the gauge field with the true gauge symmetry. Here the purpose of constructing the Wilson loop is to get a gauge invariant result for each diagram, so the perturbation result is easier to see and to compare with the open string case. 
%%%$F_{\mu \nu} = \partial_\mu A_\nv - \partial_\nv A_\mu -i [A_\mu, A_\nu]$

To describe this gauge symmetry geometrically, firstly we need to build the
Wilson line for an infinitesimal distance, then extend it to a finite length
via path ordering and finally obtain the Wilson loop which is gauge invariant.
See Peskin and Schroeder~\cite[Chapter~15]{Peskin:1995ev} for the case of
ordinary quantum field theory. 

Here we define the Wilson line for an infinitesimal separation $\epsilon$
in the same way as the open string case eq.~\eqref{eq:openWilsonLine}
\begin{align}
  \label{eq:phaseFactor}
  V[z_1+\epsilon, z_1]  &\coloneqq   \exp\{ i \int_{z_1}^{z_1+\epsilon} dz [\partial_z X^\mu A_\mu -
      \frac{1}{2} F_{\nu_1 \nu_2} \phi^{\nu_1} \phi^{\nu_2}]\},
\end{align}
except that here $\epsilon$ is on the complex plane while in the open string case it is on the boundary. 
The parametrization of the Wilson line is actually $V[z_2, z_1] = V[X, \phi, z_2, z_1]$, but for
convenience we will not explicitly distinguish them and will use whatever is convenient.
Under the gauge transformation, this Wilson line transforms as
%%Using series expansion, and assuming slow field variation of $A_\mu(X)$.
\begin{equation}
  \label{eq:phaseFactorTransf}
  V[z_1 + \epsilon, z_1] \rightarrow U[z_1 + \epsilon]  V[z_1 + \epsilon, z_1] U[z_1]^\dagger,
\end{equation}
which is exactly the property we expect for the Wilson line. Now we define the Wilson loop by taking the trace for a path ordered loop of this Wilson line
\begin{equation}
  \label{eq:heteWilLoop}
  W[X,\phi, C] \coloneqq \Tr V[X,\phi,C, z, z],
\end{equation}
where $C$ is a loop starting from $z$ and ending at $z$ and the dependence on
the bosonic string $X$ and the fermionic string $\phi$ is written explicitly. 
Compared with the Wilson loop of the open string in
section~\ref{sec:openWilsonLoop}, the only difference  is the contour on
the world-sheet of the integral. For the open string, the contour is the 1-dim
boundary. For the heterotic string, it can be any loop on the
complex plane. 

This definition of the Wilson line is simply an analog of the open string case and is not enough to justify itself. In the case of open string, the key property is that the Wilson line is the exact propagator of the fermion field eq.~\eqref{eq:openPhasePropagator}. To justify the definition for the heterotic case, we need to prove an analog of this property. To do this, we firstly need to investigate the background field expansion of the Wilson loop. 

\subsubsection{The functional variation of the Wilson loop}
\label{sec:heteBGFConv}

The perturbation calculation using the Wilson loop  is usually
done in the background field expansion. 
The background field expansion of the  bosonic string is $X \rightarrow X+\delta
X$, where $\delta X$ is a quantum field. For the fermionic string, it  is taken as a quantum field itself  $\phi=0+\phi$. 
In this way, the fermionic string contribution is an ordinary derivative 
\[V[X+\delta X,\phi,z_0, z_0] =  V[X +\delta X, z_0, z_0] +  \oint dz_1 V[X,z_0,z_1]\{
  - \frac{i}{2} F_{\nu_1 \nu_2}(X(z_1)) \phi^{\nu_2} \phi^{\nu_1} \} V[X, z_1, z_0]. \]
On the other hand, the functional derivative of the bosonic string contribution is highly
nontrivial.  We will follow  the reference~\cite{WilsonLoopExpansion} to
explain how to compute $V[X +\delta X, z_0, z_0]$. 

\begin{figure}
  \centering
  \includegraphics[width=10cm]{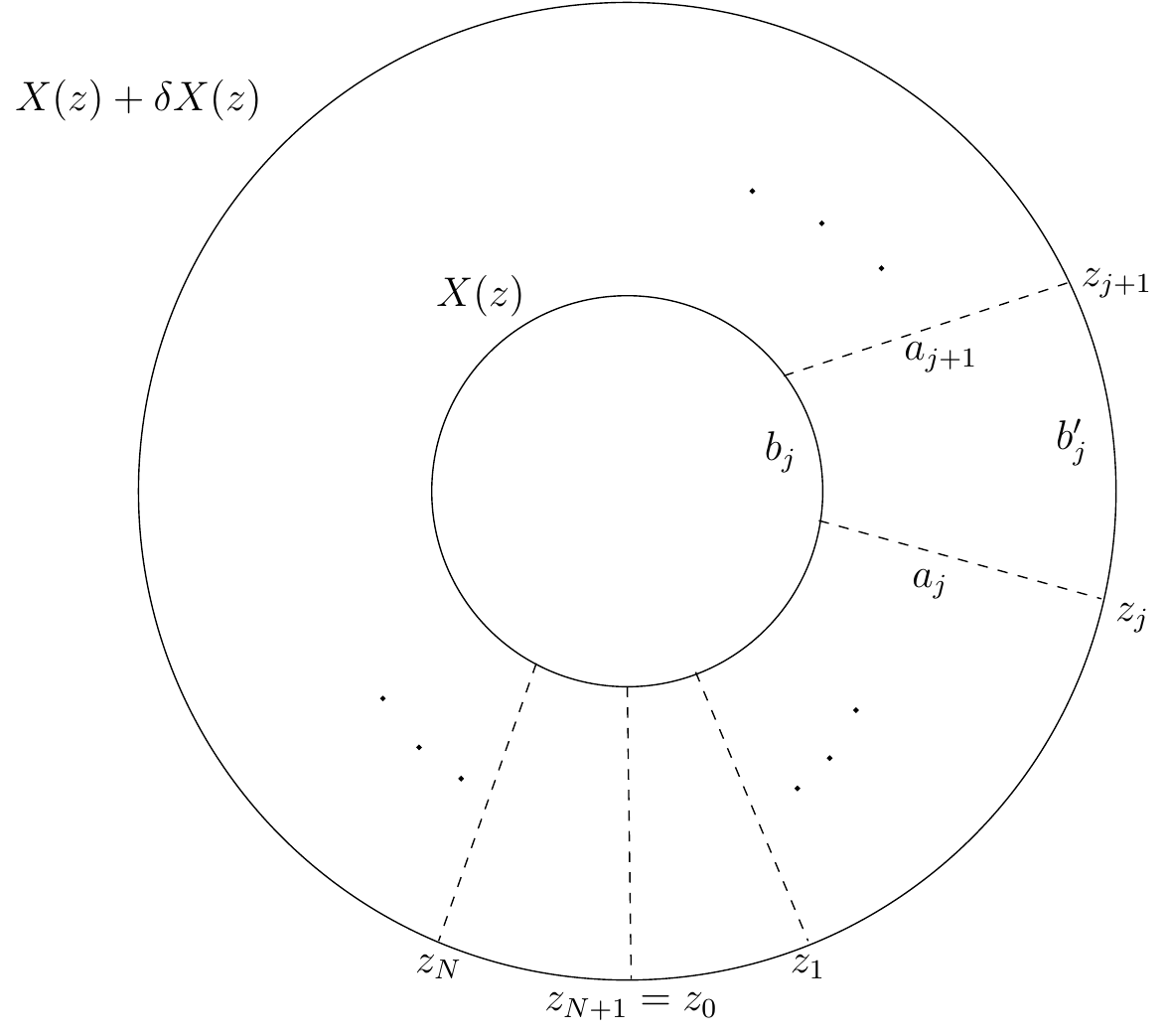}
  \caption{The discretized contour of the Wilson loop in spacetime. The $z_j$ are the holomorphic
    coordinate on the complex plane parameterizing the string $X$. }
  \label{fig:discretized_loop}
\end{figure}

For the Wilson loop, the background field expansion $X+\delta
X$ is just a variation of the contour of the loop integral and the world-sheet
coordinate serves as a parametrization of this contour. To do the functional
derivative, we discretized the contour by $z_{j+1} - z_j = \epsilon$ and compute
the variation, then take the limit $\epsilon \to 0$. The discretized loop is shown in
figure~\ref{fig:discretized_loop}. 
Then  we define the following quantities
\begin{align}
  \label{eq:heteLoopAB}
  a_j &= 1 + i A_\mu(X(z_j)) \delta X^\mu(z_j)  \nonumber \\
  a_{j+1} &= 1 + i A_\mu(X(z_{j+1})) \delta X^\mu(z_{j+1}) \nonumber \\
  b_j &= 1 + i A_\mu(X(z_j)) (X^\mu(z_{j+1}) - X^\mu(z_j)) \nonumber \\
  b'_j &= 1 + i A_\mu(X(z_j)+\delta X(z_j))(X^\mu(z_{j+1}) +\delta X^\mu(z_{j+1}) -
           X^\mu(z_j) - \delta X^\mu(z_j)), 
\end{align}
where $b_j$ represents the infinitesimal segment of $V[X +\delta X, z_0, z_0]$ and $a_j$
characterizes the change of this infinitesimal segment under the variation $X
\rightarrow X+\delta X$.
After a few algebra and just keeping the leading order of variation, we have the
following result
\begin{equation}
  \label{eq:heteLoopABAB}
  a^{-1}_j b'^{-1}_j a_{j+1} b_j = 1 - i F_{\mu\nu}(X(z_j))(X^\nu(z_{j+1}) - X^\nu(z_j)) \delta X^\mu(z_j).
\end{equation}
Taking its inverse, we get the infinitesimal segment of the Wilson loop after the variation
\[  b'_j =  a_{j+1} b_j (1 + i F_{\mu\nu}(X(z_j))(X^\nu(z_{j+1}) - X^\nu(z_j))
  \delta X^\mu(z_j)) a^{-1}_j. \]
Now the whole Wilson line after the variation is
\begin{align}
  \label{eq:heteWilDesWho}
  \prod_{j=0}^{N} b'_j &= b'_{N}  b'_{N-1}\ldots b'_{0}   \nonumber \\
  &= a_{N+1} b_N (1 + i F_{\mu\nu}(X(z_{N}))(X^\nu(z_{N+1}) - X^\nu(z_N))
      \delta X^\mu(z_N)) a^{-1}_N\nonumber \\
  &{} \times a_N b_{N-1} (1 + i F_{\mu\nu}(X(z_{N-1}))(X^\nu(z_{N}) - X^\nu(z_{N-1}))
       \delta X^\mu(z_{N-1}))  a^{-1}_{N-1} \nonumber \\
  &{} \ldots  a_1 b_0  (1 + i F_{\mu\nu}(X(z_{0}))(X^\nu(z_{1}) - X^\nu(z_0))
      \delta X^\mu(z_0)) a^{-1}_0 \nonumber \\
   &= a_{0} b_{N} b_{N-1} \ldots  b_{0}  a^{-1}_0 \nonumber\\
  &{} + \sum_{j=0}^{N} \epsilon a_0 b_N
      \ldots b_{j+1}  [ i F_{\mu\nu}(X(z_{j})) \frac{X^\nu(z_{j+1}) - X^\nu(z_j)}{z_{j+1}-z_j}
      \delta X^\mu(z_j)] b_{j} \ldots b_0 a^{-1}_0. 
\end{align}
Take the continuous limit $\epsilon \rightarrow 0$, we obtain the functional
variation of the bosonic part of the Wilson line
\begin{align}
  \label{eq:heteFuncWil}
  V[X+\delta X,z_0, z_0] &= a_0  V[X,z_0, z_0] a^{-1}_0 + a_0 \oint dz V[X,z_0,
  z] (i F_{\mu\nu}(X(z)) \partial X^\nu(z)  \delta X^\mu(z) ) V[X,z, z_0]
                             a^{-1}_0 \nonumber \\
  &= V[X,z_0, z_0] + i (A_\mu(X(z_0)) V[X,z_0,z_0]  -  V[X,z_0,z_0]
      A_\mu(X(z_0))) \delta X^\mu(z_0) \nonumber \\
  &{} +  \oint dz V[X,z_0,
  z] (i \partial X^\nu(z) F_{\mu\nu}(X(z)) \delta X^\mu(z) ) V[X,z, z_0].
\end{align}
 Combine with the fermionic part, we get the complete variation  
\begin{align}
  \label{eq:functionalDePhaseLoop}
  V[X+\delta X,\phi,z_0, z_0] &= V[X, z_0, z_0] + i \delta X^\mu(z_0) A_\mu V[X, z_0, z_0] 
                             - i V[X, z_0, z_0] A_\mu \delta  X^\mu(z_0) \nonumber \\
  &{}+i \oint dz_1 V[X,z_0,z_1]\{ \partial_z X^\nu(z_1) F_{\mu\nu}(X(z_1))
      \delta X^\mu(z_1)  \nonumber \\
  &{}- \frac{1}{2} F_{\nu_1 \nu_2}(X(z_1)) \phi^{\nu_1} \phi^{\nu_2} \} V[X, z_1, z_0].
\end{align}
Taking its trace we get the function variation of the Wilson loop
\begin{align}
  \label{eq:functionalDePhaseLoop1}
  W[X+\delta X,\phi] &= W[X] + i \oint dz_1 \Tr \{V[X](\partial_z X^\nu(z_1) F_{\mu
                    \nu}(X(z_1)) \delta X^\mu(z_1) \nonumber \\
  &{}-      \frac{1}{2} F_{\nu_1 \nu_2}(X(z_1)) \phi^{\nu_1} \phi^{\nu_2}) \},
\end{align}
The first part in the integrand is in the contribution from the bosonic string
$X$ as in reference~\cite{WilsonLoopExpansion} and the second part is the
contribution from the fermionic string $\phi$.  From this functional variation, we can obtain the  background field expansion of the Wilson loop 
\begin{align}
  \label{eq:heteBGEWilson}
  W[X+\xi,\phi] &=  i \oint dz \Tr V[X] \left\{ \partial_z X^\mu A_\mu  + \partial X^\nu [F_{
                    \mu_1\nu} \xi^{\mu_1}+ \sum_{n=2}\frac{1}{n!} D_{\mu_n}\ldots
       D_{\mu_2} F_{ \mu_1\nu} \xi^{\mu_1}\ldots \xi^{\mu_n} ]\right. \nonumber \\
  &{} \left.+[ \frac{1}{2} F_{ \mu_1\mu_2} \xi^{\mu_1} \partial \xi^{\mu_2} +\sum_{n=3} \frac{n-1}{n!} D_{\mu_{n-1}}\ldots
       D_{\mu_2} F_{ \mu_1\mu_n} \xi^{\mu_1}\ldots \xi^{\mu_{n-1}} \partial
       \xi^{\mu_n}] \right. \nonumber\\
  &{} \left.-\frac{1}{2} [F_{\nu_1 \nu_2} \phi^{\nu_1}  \phi^{\nu_2} +
       \sum_{n=1}\frac{1}{n!}  D_{\mu_n}\ldots D_{\mu_1}F_{\nu_1 \nu_2}(X)
       \phi^{\nu_1}\phi^{\nu_2}  \xi^{\mu_1}\ldots \xi^{\mu_n} ] \right\}.
\end{align}
This result is the same as eq.~\eqref{eq:openBGEWilson}, except the difference between parameters $z$ and $\tau$. Now the Wilson loop between the open string case and the heterotic case corresponds to each other very well. In section~\ref{sec:geoWilson}, we will construct the action of the gauge physics using the Wilson loop for the heterotic sigma model, so the open string sigma model and the heterotic sigma model can corresponds to each other in the level of action.

\subsubsection{The exact propagator of psi}
\label{sec:HetePhasePropagator}

Now let's prove that the Wilson loop is the exact propagator of the fermion field $\psi$.
Firstly, from the path ordering of the Wilson loop, we would have the following differential equation
\begin{equation}
  \label{eq:dePhaseFactor}
  \frac{d}{dz_2} V[z_2, z_1] = i (\partial_{z_2} X^\mu A_\mu(X(z_2)) -
  \frac{1}{2} F_{\nu_1 \nu_2}(X(z_2)) \phi^{\nu_1} \phi^{\nu_2}) V[z_2, z_1], 
\end{equation}
which is just an analog of the differential equation of the time evolution
operator in quantum field theory. Integrating out this differential equation gives us the path ordered
Wilson line for a curve of finite length. 
This equation is essentially equivalent to the following variation
\begin{align}
  \label{eq:heteIntgDiff}
  V[z_2+\epsilon_2, z_1] = V[z_2 , z_1] + \epsilon_2 i  (\partial_{z_2} X^\mu A_\mu(z_2) -
  \frac{1}{2} F_{\nu_1 \nu_2}(z_2) \phi^{\nu_1} \phi^{\nu_2}) V[z_2, z_1]. 
\end{align}
Secondly, we can obtained this variation in a different way using eq.~\eqref{eq:functionalDePhaseLoop}
\begin{align}
  \label{eq:heteFuncDerPhase}
   V[z_2+\epsilon_2, z_1] &=  V[X + \delta X,\phi+\delta \phi, z_2+\epsilon_2,
                              z_1]  \nonumber \\
                          &= V[X, \phi, z_2, z_1] + \frac{\partial
                              V[z_2, z_1]}{\partial z_2} \epsilon_2  \nonumber \\
  &{} + i A_\mu(z_2) \delta X(z_2) V[z_2, z_1] - i V[z_2, z_1] A_\mu(z_1) \delta
      X(z_1)  \nonumber \\
  &{}+ i \oint dz V[z_2, z](\partial X^\nu(z) F_{\mu\nu} \delta
      X^\mu(z) -  F_{\nu_1 \nu_2} \phi^{\nu_1} \delta\phi^{\nu_2})V[z,z_1].
\end{align}
Use $\delta X(z) = \partial X(z) \delta (z-z_2) \epsilon_2$ and $\delta \phi
=\partial_z \phi \delta(z-z_2) \epsilon_2 =0$, the above equation becomes
%%%classical e.o.m. for $\phi$
\begin{align}
  \label{eq:heteFuncVarPha}
   V[z_2+\epsilon_2, z_1] &= V[X, \phi, z_2, z_1] + \frac{\partial
                              V[z_2, z_1]}{\partial z_2}  \epsilon_2
  + i A_\mu(z_2)\partial X^\mu(z_2) \delta(0) V[z_2, z_1] \epsilon_2 \nonumber \\
  &{}- i V[z_2, z_1] A_\mu(z_1)\partial X^\mu(z_1)
       \delta (z_2 - z_1) \epsilon_2  + i\partial
       X^\mu(z_2) \partial X^\nu(z_2) F_{\mu\nu} V[z_2, z_1] \epsilon_2\nonumber \\
  &= V[X, \phi, z_2, z_1] + \frac{\partial
                              V[z_2, z_1]}{\partial z_2}  \epsilon_2  + i
      A_\mu(z_2) \partial X^\mu(z_2) \delta(0) V[z_2, z_1] \epsilon_2 \nonumber \\ 
    &{}- i V[z_2, z_1] A_\mu(z_1)\partial X^\mu(z_1) \delta (z_2 - z_1) \epsilon_2.
\end{align}
Combining the two different ways of doing the variation,
eq.~\eqref{eq:heteIntgDiff} and eq.~\eqref{eq:heteFuncVarPha}, we get
\begin{align}
  \label{eq:heteIntOut}
 &[ \frac{\partial}{\partial z_2} - i (\partial_{z_2} X^\mu A_\mu(z_2) -
  \frac{1}{2} F_{\nu_1 \nu_2}(z_2) \phi^{\nu_1} \phi^{\nu_2}) ] V[z_2, z_1]
                                              \nonumber \\
  &= -i A_\mu(z_2) \partial X^\mu(z_2)  V[z_2, z_1] \delta(0) + i V[z_2, z_1] A_\mu(z_1) \partial X^\mu(z_1) \delta (z_2 - z_1). 
\end{align}
Except the $\delta(0)$ term, the Wilson line $V[z_2, z_1]$ is the inverse of the differential operator 
\begin{equation}
\frac{\partial}{\partial z_2} - i (\partial_{z_2} X^\mu A_\mu(z_2) -
  \frac{1}{2} F_{\nu_1 \nu_2}(z_2) \phi^{\nu_1} \phi^{\nu_2}).
\end{equation}
Compared with eq.~\eqref{eq:heteFermPart}, we see that $V[z_2,z_1]$ is the
exact propagator of the fermion field $\psi$, 
\begin{align}
  \label{eq:heteConnectionWilFer}
  V[z_2, z_1] = \langle \psi(z_2) \psi(z_1) \rangle_{L_E^f}. 
\end{align}
For the  $\delta(0)$ term,  it can be incorporated into the
normalization of the partition function of $\psi$, so we can just throw away this infinity term from the partition function. 

Now  this definition of the Wilson loop for the heterotic sigma model is justified. The relation between the Wilson loop and the fermionic
representation for the heterotic sigma model, is exactly the same as that
relation for the open string sigma model. The contour integral of the Wilson
loop is equivalent to the ordinary perturbation in terms of the fermion field
$\psi$. For the heterotic string, this relation is highly nontrivial, because
the fermion field $\psi$ lives on the whole complex plane.   For the open
string, this relation is a trivial one, because its fermionic field $\psi$ just
lives on the boundary and its propagator is just the Heaviside step function.

\subsection{Geometry of the Wilson loop}
\label{sec:geoWilson}
The Wilson loop is a geometrical object that carry the gauge physics. 
To build up the action of the heterotic string sigma model using the Wilson loop, we need to explore its geometry.
Firstly we will  look at how we arrive at the classical Yang-Mills action using the
Wilson loop and this will serve as a protocol.  Then we will discuss the open string case and the 
heterotic string case. In all of these theories, the gauge invariant classical action is 
obtained from a sum over the loop contours, which is equivalent to a sum of all the
gauge content over the spacetime. 

\subsubsection{Yang-Mills case}

For the Yang-Mills case, we will use the lattice theory as a convenient
illustration. In the lattice theory, the sum of Wilson loop over all the loops
will generate the Yang-Mills action, as shown in the 
following equation (See Srednicki~\cite[Chapter~82]{Srednicki:2007qs} for details)
\begin{equation}
  \label{eq:wilsonActionYM}
  S \approx \sum_{loops} W[\textrm{plaquette}],
\end{equation}
where $W[\textrm{plaquette}]$ is the Wilson loop associated with a specific plaquette.
\emph{This equivalence of the gauge 
invariant action and the sum over all the Wilson loops comes from the geometric
nature of the Wilson loop}.

The line integral of the gauge field $A_\mu$ in the Wilson loop is connected
with  the area integral of the field strength $F_{\mu\nu}$ via
\begin{equation}
  \label{eq:heteWilsonStokes}
\Tr \mathcal{P} \exp\{ i \oint_C dx^\mu A_\mu(x)\} = \Tr \mathcal{P} \exp\{ i \int_\Sigma d\sigma^{\mu\nu}(x) V[x_0,x] F_{\mu\nu}(x)V[x,x_0]\},
\end{equation}
where $\sigma^{\mu\nu}(x)$ is the area element on the surface $\Sigma$ bounded by
the closed loop $C$ and $V[x_0,x] = \mathcal{P} \exp\{ i \int_x^{x_0} dy^\mu A_\mu(y)\}$. 
In the abelian case, this is simply the Stokes's theorem. In the nonabelian case,
this is called the nonabelian Stokes's theorem and is highly
nontrivial~\cite{nonabelianStokes1,nonabelianStokes2,nonabelianStokes3}.

\begin{figure}
  \centering
  \includegraphics[width=10cm]{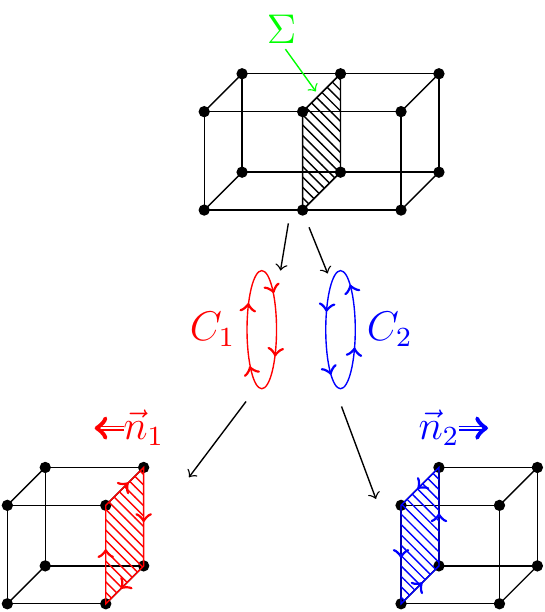}
  \caption{The sum of the Wilson loop over both the directions. 
    In the lattice, each plaquette $\Sigma$  has two loops of opposite directions $C_1$
    and $C_2$. By the Stokes's theorem, the line integrals over $C_1$ and $C_2$
    are  connected with  area integrals over the two area (faces) that have opposite
    normal directions $\vec{n}_1$ and $\vec{n}_2$. In this way, the Wilson loop
    over $C_1$ and $C_2$  are connected with the gauge contribution to the left
    cube and the right cube respectively.  }
  \label{fig_3D_lattice}
\end{figure}

Now let's go to a lattice theory to see how this Stokes's theorem leads to the
sum over loops. For simplicity (to be able to draw the figure), let's assume a
3D spacetime lattice. And we choose the right-hand rule to associate the direction of the
area with the direction of the loop. The gauge contribution to the
Yang-Mills action should be a volume integral over the 3D spacetime.
However, on the lattice,  the gauge content is only defined on the 1D loops (the
boundary of the plaquette). By
the Stokes's theorem, we extend the gauge content from the 1D loop to the 2D
area (plaquette) bounded by the loop. In this way, the volume integral of the gauge
content over the unit cube becomes the sum of the area integral of the 
gauge content over all the plaquettes of the cube.  Let's look at figure~\ref{fig_3D_lattice}
for illustration.
The gauge content $\exp[ i \oint_{C_{1,2}} dX^\mu A_\mu]$ is defined by the
Wilson loop integral over the boundary of the plaquette $\Sigma$, where there
are two opposite directions $C_1$ and $C_2$ for 
the 1D loop. By Stokes's theorem, the gauge content is extended to two area
integrals over the plaquette (face of the cube) $\exp[ i \int_{\vec{n}_{1,2}}
d\sigma^{\mu\nu} F_{\mu\nu}]$, with  opposite normal directions $\vec{n}_1$ and 
$\vec{n}_2$ of the area. The area integral with normal direction 
$\vec{n}_1$ is associated with the gauge content of the left unit cube and the area integral with normal direction 
$\vec{n}_2$ is associated with the gauge content of the right unit cube .

In this way, the classical action $S(\Sigma)$ of this plaquette $\Sigma$, which
is just the sum over both the directions of the Wilson loop $W[C_{1,2}]$,  turns out to  be a
sum of the gauge physics from the left cube and the right cube, which are all
the unit cubes that are adjacent to the plaquette
\begin{align}
  \label{eq:wilSumYM}
  S(\Sigma) &= \sum_{j=1}^{2}\exp[ i \oint_{C_{1,2}} dX^\mu A_\mu] = \sum_{j=1}^{2} \exp[ i \int_{\vec{n}_{1,2}}
                                                          d\sigma^{\mu\nu}
        F_{\mu\nu}]  \nonumber \\
  &= \textrm{gauge content from the left cube} + \textrm{gauge content from the
      right cube} \nonumber \\
  &= \textrm{sum of all the gauge content around $\Sigma$}.
\end{align}
So the lattice Yang-Mills action is just the sum of $\exp[ i \int_{\vec{n}}
d\sigma^{\mu\nu} F_{\mu\nu}]$ over all the elementary oriented areas of the
spacetime lattice. Going to the continuum limit by taking the infinitesimal
lattice spacing, the classical Yang-Mills action is equivalent to the sum of the
gauge physics  $\exp[ i \int_{\vec{n}} d\sigma^{\mu\nu} F_{\mu\nu}]$  over all
the infinitesimal oriented areas of the whole spacetime.

We should keep in mind that in 3D and higher dimensional spacetime, this geometric nature
of the Wilson loop only serves as an intuitive picture of the gauge physics,
because this sum over plaquettes can only be done in the lattice approximation
rather than the continuum limit. However, for strings, this geometric picture is
a practical method to do the calculation, because the 2D nature of world-sheet of the string,
as will be explored in the following.

\subsubsection{Open string case}

For the open string sigma model, the functional derivative of the Wilson
loop eq.~\eqref{eq:openBGEWilson} is  
\begin{equation}
  \label{eq:openWilsonStokes}
    W[X+\delta X,\phi] = W[X] + i \oint d\tau \Tr V[X] \{  \partial X^\nu F_{
                    \mu_1\nu} \delta X^{\mu_1}  - \frac{1}{2} F_{\nu_1 \nu_2}
                     \phi^{\nu_1}  \phi^{\nu_2}  \} + O((\delta X)^2).
\end{equation}
If we just look at the bosonic string part (set $\phi=0$), this equation is just
the nonabelian Stokes's theorem investigated
in~\cite{nonabelianStokes2,nonabelianStokes3,Dorn:1986dt}. The area element is
$\delta\sigma_{bosonic}^{\mu\nu}=  \delta\tau \partial X^\nu \delta X^\mu$ and the functional variation is an area integral of $F_{\mu\nu}$. We will obtain the nonabelian
Stokes's theorem of the bosonic open string~\cite{Dorn:1986dt, nonabelianStokes2}
\begin{equation}
  \label{eq:openWilsStokesBos}
\Tr \mathcal{P} \exp\{ i\oint d\tau  \partial X^\mu(\tau) A_{\mu}(X)\}  = \Tr \mathcal{P} \exp\{ i \int d\sigma_{bosonic}^{\mu\nu}(X) V[X_0,X] F_{\mu\nu}(X(\tau))V[X,X_0]\},
\end{equation}
where $V[X_0(\tau_0),X(\tau)] = \mathcal{P} \exp\{ i\int_\tau^{\tau_0} d\tau'  \partial
X^\mu(\tau') A_{\mu}(X)\}$. 

Now let's turn on the fermionic string $\phi \neq 0$ and treat $\phi$ itself as the variation (like $\delta X$). By analog of the bosonic area element, we define the area element of
the fermionic string to be $\delta\sigma_{fermionic}^{\mu\nu}=\delta\tau\phi^{\mu}\phi^{\nu}$ in
the Grassmann space. Now the fermionic part of the functional
derivative also becomes an area integral of $F_{\mu\nu}$.  Like the bosonic case, we can integrate out the functional
derivative and obtain a fermionic contribution to the nonabelian Stokes's
theorem. So we obtain a generalization of the nonabelian Stokes's theorem to the superstring 
\begin{align}
  \label{eq:openWilsStokesGene}
  W[C]&=\Tr \mathcal{P} \exp\{ i\oint d\tau [   \partial X^\mu(\tau) A_{\mu}(X)
          - \frac{1}{2} F_{\nu_1 \nu_2}(X) \phi^{\nu_1}  \phi^{\nu_2} ] \}
          \nonumber \\ 
      &= \Tr \mathcal{P} \exp\{ i \int [d\sigma_{bosonic}^{\mu\nu}
          -\frac{1}{2} d\sigma_{fermionic}^{\mu\nu}] V[X_0,X]
          F_{\mu\nu}(X)V[X,X_0] \},
\end{align}
where $V[X_0(\tau_0),X(\tau)] = \mathcal{P} \exp\{ i\int_\tau^{\tau_0} d\tau' [ \partial X^\mu(\tau') A_{\mu}(X) - \frac{1}{2} F_{\nu_1 \nu_2}(X) \phi^{\nu_1}  \phi^{\nu_2} ] \}$. 

\begin{figure}
  \centering
  \includegraphics[width=8cm]{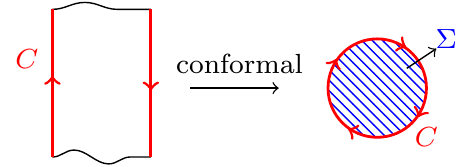}
  \caption{The geometry of the Wilson loop of the open string. On the left is
    the open string in spacetime. On the right is its conformal transformation
    into unit disk. The loop is $C$ and the area bounded is $\Sigma$.  }
  \label{fig_open_wilson}
\end{figure}

From this generalized nonabelian Stokes's theorem, we can see the
geometry of the Wilson loop in the open string case and then obtain the
classical action from a sum over loops like the Yang-Mills case. Let's look at
figure~\ref{fig_open_wilson} for illustration.
For open string, the gauge field only lives on the boundary $C$ of the string
via the Chan-Paton factors.
So the only loop we have is the boundary itself. After conformal transformation
into the unit disk, the area $\Sigma$ bounded by $C$ is the disk itself. So by
the Stokes's theorem, the single loop $C$ would contain all the contribution of
the gauge field of the open string. This explains the fact that in the open string
sigma model we only use a single Wilson loop eq.~\eqref{eq:openWilsonLoop}
without any sum over loops, because this already includes all the gauge contributions of the string.

\subsubsection{Heterotic string case}

For the heterotic string, the generalized nonabelian Stokes's theorem can be
obtained straightforwardly following the discussion of the open string case
\begin{align}
  \label{eq:heteStoGen}
    W[C]&=\Tr \mathcal{P} \exp\{ i\oint dz [   \partial X^\mu(z) A_{\mu}(X)
          - \frac{1}{2} F_{\nu_1 \nu_2}(X) \phi^{\nu_1}  \phi^{\nu_2} ] \}
          \nonumber \\ 
      &= \Tr \mathcal{P} \exp\{i \int [d\sigma_{bosonic}^{\mu\nu}
          -\frac{1}{2} d\sigma_{fermionic}^{\mu\nu}] V[X_0,X]
          F_{\mu\nu}(X)V[X,X_0] \}
\end{align}
where $V[X_0(z_0),X(z)]= \mathcal{P} \exp\{ i \int_z^{z_0} dz' [\partial
X^\mu(z') A_\mu - \frac{1}{2} F_{\nu_1 \nu_2} \phi^{\nu_1} \phi^{\nu_2}]\} $. 
This is nearly the same as the open string one
eq.~\eqref{eq:openWilsStokesGene}, except that here the parametrization is $z$.

\begin{figure}
  \centering
  \includegraphics[width=12cm]{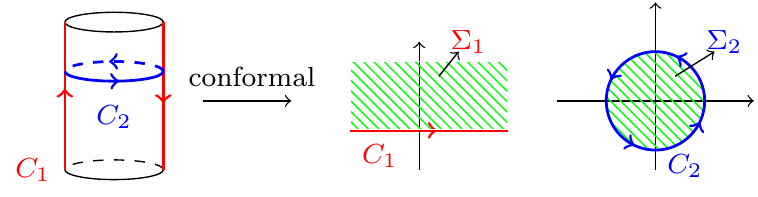}
  \caption{The geometry of the Wilson loop of the heterotic string. On the left is
    the heterotic string in spacetime. There are two types of loops, the
    longitudinal one $C_1$ and the transversal one $C_2$.  On the right are the
    conformal transformation of the closed string, where $C_1$ is transformed
    into the real axis and $C_2$ is transformed into a circle. The area bounded
    by the two loops are $\Sigma_1$ and $\Sigma_2$ respectively, as can be
    distinguished by their color. }
  \label{fig_hete_wilson}
\end{figure}

Let's look at figure~\ref{fig_hete_wilson} for the geometry of the Wilson loop
of the heterotic string. 
There are two types of loops on the closed string, the longitudinal one $C_1$
and the transversal one $C_2$. The conformal transformation maps the closed string into
the whole complex plane and $C_{1,2}$ are mapped into the real axis and the
circle respectively. Let's focus on $C_1$ first. The area bounded by $C_1$ is
the upper half-plane $\Sigma_1$. By the nonabelian Stokes's theorem, the Wilson loop of
$C_1$ would give the gauge physics of the upper half-plane. If we revert the
direction $-C_1$, the area bounded will be the lower half-plane and the
nonabelian Stokes's theorem would give the gauge physics of the lower
half-plane. So if we sum the Wilson loop over the loop  $C_1$ and $-C_1$, we
will have the gauge physics of the whole complex plane, thus of the whole closed
string. Now look at $C_2$. It is straightforward to see that the sum of the
Wilson loop over $C_2$ and $-C_2$ will also give the gauge physics of the whole
complex plane.

This result can be generalized to an arbitrary loop $C$. Because of the 2D
nature of the closed string, the two areas bounded by $C$ and $-C$ are
complementary and their sum is the whole complex plane. So by nonabelian
Stokes's theorem, we arrive at the following proposal for the Wilson loop
approach of the heterotic string
\begin{align}
  \label{eq:effectiveActionCalculation1}
  e^{-S_{eff}[F_{\mu\nu}]} &= \sum_{\pm C} \int D\phi DX e^{-S_E[X,A,\phi, C]}
                 \nonumber\\
             &= \sum_{\pm C} \int D\phi DX \Tr \mathcal{P} \exp\{-
                 \frac{1}{2 \pi \alpha'} \int d^2z [  \partial X^{\mu
                 }\bar{\partial}X_{\mu } + \phi ^{\mu } \bar{\partial }\phi _{\mu }]
                 \nonumber \\
  &{} + i\oint_C dz [   \partial X^\mu(z) A_{\mu}(X)  - \frac{1}{2}  F_{\nu_1
       \nu_2}(X) \phi^{\nu_2}  \phi^{\nu_1} ] \}.   
\end{align}
This is similar to the open string case
eq.~\eqref{eq:openWilsonLoop}, except that now we have to sum over
two directions of the contour. So the background field expansion of this action parallels that of the open string case, just replace $\tau$ in eq.~\eqref{eq:openBGEWilson} with $z$. 
Unlike the Yang-Mills case where the sum over Wilson loops is only calculable in
lattice theory, here for the heterotic string, the sum over Wilson loops is
practical: pick an arbitrary loop $C$, sum the Wilson loop over both
directions $\pm C$, then the result is the classical action of 
the gauge field in spacetime.

\subsection{Path ordering and contour direction}
\label{sec:pathOrder}

Since there are two directions of the contour, we need to give a comment about its relation with the path ordering of the gauge factors. 
We will take the convention of distinguishing the path ordering of the
gauge field and the direction of the contour, i.e., we treat them as two different
kinds of ordering. Firstly, we define the direction $C_+$ and $C_-$ for a contour loop. Since we are using the upper half plane for the open string world-sheet, we define $C_+$ to be from $-\infty$ to
$\infty$ on the real axis and $C_-$ is just its inverse. 
Then, for a given vertex structure of a Feynman diagram, the gauge factors of the vertices are defined to be along the direction of $C_+$. 
Finally, when calculate this Feynman diagram, we just compute the integrand along $C_+$ in the open string case, and compute the integrand along both $C_+$ and $C_-$ in the heterotic string case.  
By this convention, we can just focus  on the computation of the integrand, and leave the vertex structure of gauge factors aside. 

\section{Single-valued map}
\label{sec:singleValuedMap}

Now we can explore the sv-map using the Wilson loop representation for both the open string and the heterotic string. We will show how sv-map arise in  three loop and four loop level for $\zeta_2$ and $\zeta_3$ respectively.
 Our purpose is to find the mechanism of the sv-map, rather than compute the complete beta function. 
 So instead of pursuing the complete renormalization, we only compute the single pole (single logarithmic divergence)  and will show that for a given Feynman diagram of single-trace terms,  the single poles of the open and the heterotic string integrals satisfy the sv-map. 
We will just focus on  bosonic loops and 
compute a few representative diagrams of three loop and four loop, and show that  sv-map connects the open case and the heterotic case in each diagram.  (Representative here means they corresponds to permutations of the vertex structure at a given loop level). 

For the computation we use the following setup (this is a recall of what we get)
\begin{align}
  \label{eq:openHeteWils}
  \textrm{open:} &=  \int D\phi DX  \Tr \mathcal{P} \exp \{ - \frac{1}{4 \pi \alpha'} \int d^2z (\partial X^\mu
               \bar{\partial} X_\mu + \Phi^\mu \bar{\partial} \Phi_\mu +
               \tilde{\Phi}^\mu \partial \tilde{\Phi}_\mu ) \nonumber\\
                   &{}  + i\oint_{C_+} d t [   \partial X^\mu(t) A_{\mu}(X)
          - \frac{1}{2} F_{\nu_1 \nu_2}(X) \phi^{\nu_1}  \phi^{\nu_2} ]  \}
                    \nonumber\\
\textrm{hete:} &= \sum_{C_+, C_-} \int D\phi DX \Tr \mathcal{P} \exp\{-
                 \frac{2}{4 \pi \alpha'} \int d^2z [  \partial X^{\mu
                 }\bar{\partial}X_{\mu } + \phi ^{\mu } \bar{\partial }\phi _{\mu }]
                 \nonumber \\
  &{} + i\oint_C dz [   \partial X^\mu(z) A_{\mu}(X)  - \frac{1}{2}  F_{\nu_1
       \nu_2}(X) \phi^{\nu_1}  \phi^{\nu_2} ] \},
\end{align}
where $C_+$ is the real axis, $C_-$ is its inverse and we will use variable $t$ for the real axis from now on. The background field expansion is given in eq.~\eqref{eq:openBGEWilson}. 
Both the propagators are $\langle X^\mu(t_1) X^\nu(t_2) \rangle = -\eta^{\mu\nu} \alpha' \ln{(t_1 - t_2)^2}$ on the real axis. The bosonic propagators are represented by wavy lines and the contour of loop (the real axis) is represented by a solid line. A slash on the wavy line represents a derivative of the propagator, with respect to the most close vertex coordinate. 

\subsection{The zeta(2) case}
\label{sec:zeta2Case}

\begin{figure}
\begin{center}
\includegraphics[width=10cm]{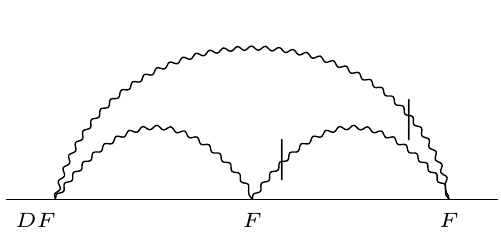}
\caption{The Feynman diagram corresponding to structure $\partial X^\nu D_{\mu_1} F_{\nu}^{~\mu_3}
F_{\mu_3}^{~\mu_4} F_{\mu_4}^{~\mu_1} $.  Wavy lines are bosonic propagators and  the solid line is the contour. The slash on the wavy line represents a derivative of the propagator, with respect to the most close vertex coordinate.}
\label{fig:open3loop}
\end{center}
\end{figure}

Now, let's look at how the sv-map of $\zeta(2)$ arises at three loop level.  The mathematical sv of $\zeta_3$ is  $sv(\zeta_2) = 0$~\cite{StiebergerSVmath14}. 
We will focus on the diagram shown in figure~\ref{fig:open3loop}.
It contributes to the sigma model an ultra-violet divergent Lagrangian term of the form $\partial X^\nu D_{\mu_1} F_{\nu}^{~\mu_3}F_{\mu_3}^{~\mu_4} F_{\mu_4}^{~\mu_1} $.
The open string integral associated to this diagram is 
\begin{align}
\label{eq:3loopCplus}
I_{3, C_+} &=(-2\alpha')^3 \int_{-\infty<t_1<t_2<t_3<\infty}dt_1 dt_2 dt_3 f(V(t_1),V(t_2),V(t_3)) \frac{\ln{t_{21}}}{t_{31} t_{32}} \nonumber\\ 
&= (-2\alpha')^3 \int_{-\infty}^{\infty} dt_3 f(V(t_3),V(t_3),V(t_3)) \int_{-\infty < t_1 < t_2 < t_3} d t_2 dt_1  \frac{\ln{t_{21}}}{t_{31} t_{32}} \nonumber \\
&\approx \int_{-\infty}^{\infty} dt_3   \int_{-\infty < t_1 < t_2 < t_3} d t_2 dt_1  \frac{\ln{t_{21}}}{t_{31} t_{32}},
\end{align}
where $t_{jk} = t_j - t_k$ and $ f(V(t_1),V(t_2),V(t_3)) =\partial X^\nu(t_1) D_{\mu_1} F_{\nu}^{~\mu_3}(t_1)
F_{\mu_3}^{~\mu_4}(t_2) F_{\mu_4}^{~\mu_1}(t_3)$ is the vertex structure.  In the last step we hide the constant factors $(-2\alpha')^3$ and the vertex structure, and just focus on the computation of the integrals. We will hide such constant factors and vertex structures in all the following computations. 

To compute it, do the following change of variables
\begin{align}
  \label{eq:3loopCplusCOV}
  w &= t_{31} \nonumber \\
  u &= \frac{t_{32}}{t_{31}} = \frac{t_{32}}{w}, \quad 0 < u < 1, 
\end{align}
which changes the original integral as following
\begin{align}
  \label{eq:3loopCplusRes}
  I_{3, C_+}  &= \int_{-\infty}^{\infty} dt_3 \int_{-\infty}^{0} dt_{13} \int_{t_{13}}^{0} dt_{23} \frac{\ln{(t_{31} - t_{32})}}{t_{31}  t_{32}} \nonumber \\
     &= \int_{-\infty}^{\infty} dt_3   \int_0^{\infty} \frac{dw}{w} \int_0^{1} du
       \frac{\ln{w}  + \ln{(1-u)}}{u} \nonumber\\
     &= \int_{-\infty}^{\infty} dt_3  \left( -\zeta_2 (\ln{\lambda} - \ln{\epsilon})\right),
\end{align}
where we use the brute force cutoff with $\epsilon$ the UV cutoff and $\lambda$ the IR cutoff. Only the single poles (single logarithmic divergence) are kept and higher order divergences are thrown away, the same to all the following computations.  

For the heterotic integral we also need the other contour
\begin{align}
\label{eq:3loopCminus}
I_{3, C_-} &= \int_{\infty}^{-\infty} dt_1   \int_{\infty > t_3 > t_2 > t_1} d t_2 dt_3  \frac{\ln{t_{21}}}{t_{31} t_{32}},
\end{align}
which is obtained in a similar manner as eq.~\eqref{eq:3loopCplus}. Notice that  the contour $C_-$ means that we start from $\infty$ and to $-\infty$ (we will not explicitly mention this from now on). 
Here we do the following change of variables
\begin{align}
  \label{eq:3loopCminusCOV}
  w &= t_{31} \nonumber \\
  u &= \frac{t_{21}}{t_{31}} = \frac{t_{21}}{w}, \quad 0< u < 1, 
\end{align}
which changes the original integral as following
\begin{align}
  \label{eq:3loopCminusRes}
  I_{3, C_-}  &= - \int_{-\infty}^{\infty} dt_1 \int_{\infty}^{0} dt_{31} \int_{t_{31}}^{0} dt_{21} \frac{\ln{t_{21}}}{t_{31}  t_{32}} \nonumber \\
     &= - \int_{-\infty}^{\infty} dt_1   \int_0^{\infty} \frac{dw}{w} \int_0^{1} du
       \frac{\ln{w}  + \ln{u}}{1-u} \nonumber\\
     &= - \int_{-\infty}^{\infty} dt_1  \left( -\zeta_2 (\ln{\lambda} - \ln{\epsilon}) \right).
\end{align}
The heterotic integral is zero, which is just a sum of $C_+$ and $C_-$ given in eq.~\eqref{eq:3loopCplusRes} and eq.~\eqref{eq:3loopCminusRes}. So we have the sv-map at three loop $sv(\zeta_2) = 0$.

\subsection{The zeta(3) case}
\label{sec:zeta3Case}

Now let's see how the sv-map of $\zeta(3)$ arises at four loop. The mathematical sv of $\zeta_3$ is  $sv(\zeta_3) = 2\zeta_3$~\cite{StiebergerSVmath14}.  We choose three representative diagrams figure.~\ref{fig_open_diagram_1}, figure.~\ref{fig_open_diagram_2} and figure.~\ref{fig_open_diagram_3}. 

\subsubsection{Case 1}
\label{sec:case_1}
\begin{figure}
  \centering
  \includegraphics[width=10cm]{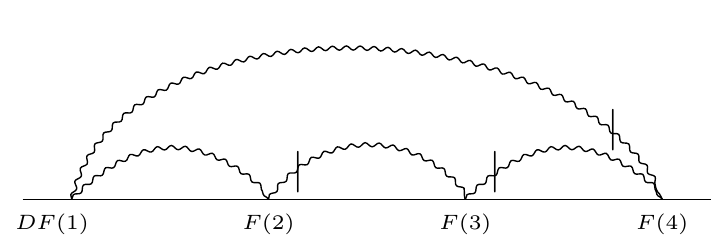}
  \caption{The Feynman diagram corresponding to structure $\partial X^\nu D_{\mu_1} F_{\nu}^{\quad\mu_3}
F_{\mu_3}^{\quad\mu_4} F_{\mu_4}^{\quad\mu_5} F_{\mu_5}^{\quad\mu_1}$. }
  \label{fig_open_diagram_1}
\end{figure}

Firstly we compute diagram of figure.~\ref{fig_open_diagram_1}, which has the vertex structure  $\partial X^\nu D_{\mu_1} F_{\nu}^{\quad\mu_3}
F_{\mu_3}^{\quad\mu_4} F_{\mu_4}^{\quad\mu_5} F_{\mu_5}^{\quad\mu_1}$.
 The
 open string integral is  
\begin{equation}\label{eq:41Cplus}
I_{41, C_+} = \int_{-\infty}^{\infty} d t_4 \int_{-\infty<t_1<t_2<t_3<t_4}dt_1 dt_2 dt_3 \frac{\ln{t_{21}}}{t_{41}
    t_{32}t_{43}}.
\end{equation}
Using the following change of variables
\begin{align}
  \label{eq:41CplusCOV}
  w &= t_{41} \nonumber \\
  v &= \frac{t_{42}}{t_{41}} = \frac{t_{42}}{w} \nonumber \\
  u &= \frac{t_{43}}{t_{42}},  0 < u, v < 1.
\end{align}
we get 
\begin{align}
  \label{eq:41CplusRes}
I_{41, C_+}   &= \int_{-\infty}^{\infty} dt_4 \int_{-\infty}^0 dt_{14} \int_{t_{14}}^0 dt_{24} \int_{t_{24}}^0 dt_{34}
       \frac{\ln{(t_{41} - t_{42})}}{t_{41}
       (t_{42} - t_{43}) t_{43}} \nonumber \\
    &= \int_{-\infty}^{\infty} dt_4   \int_0^{\infty} \frac{dw}{w} \int_{0}^{1} dv \int_0^1 du
       \frac{\ln{w}  + \ln{(1-v)}}{u v (1-u)}\nonumber\\
       &= 0.
\end{align}
There is no single poles.This is consistent with the results of~\cite{fiveAmp:effectiveAction}, where no effective action terms were found corresponding to this respective part of the beta function.

The
 heterotic string integral needs the other contour  
\begin{equation}\label{eq:41Cminus}
I_{41, C_-} = \int_{\infty}^{-\infty} d t_1 \int_{\infty>t_4>t_3>t_2>t_1}dt_4 dt_3 dt_2 \frac{\ln{t_{21}}}{t_{41}
    t_{32}t_{43}}.
\end{equation}
Using the change of variables
\begin{align}
  \label{eq:41CminusCOV}
  w &= t_{41} \nonumber \\
  v &= \frac{t_{31}}{t_{41}} = \frac{t_{31}}{w} \nonumber \\
  u &= \frac{t_{21}}{t_{31}},  0 < u, v < 1.
\end{align}
we get
\begin{align}
  \label{eq:41CminusRes}
I_{41, C_-}   &= - \int_{-\infty}^{\infty} dt_1 \int_{\infty}^0 dt_{41} \int_{t_{41}}^0 dt_{31} \int_{t_{31}}^0 dt_{21}
       \frac{\ln{(t_{21})}}{t_{41}
       (t_{31}-t_{21}) (t_{41}-t_{31})} \nonumber \\
    &= \int_{-\infty}^{\infty} dt_1   \int_0^{\infty} \frac{dw}{w} \int_{0}^{1} dv \int_0^1 du
       \frac{\ln{w}  + \ln{v} + \ln{u}}{(1-u) (1-v)} \nonumber\\
       &=  0.
\end{align}
The heterotic integral is the sum of eq.~\eqref{eq:41CplusRes} and eq.~\eqref{eq:41CminusRes}, while the open string integral is just eq.~\eqref{eq:41CplusRes}. So  we see that  $sv(0) = 0$. 

\subsubsection{Case 2}
\label{sec:case_2}
\begin{figure}
  \centering
  \includegraphics[width=10cm]{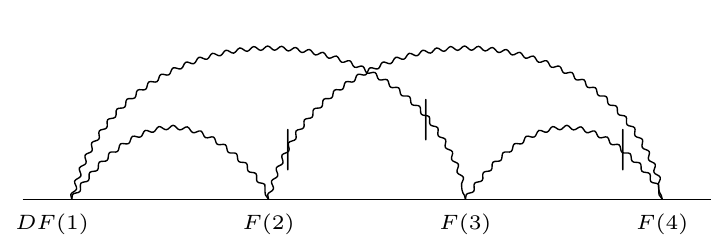}
  \caption{The Feynman diagram corresponding to structure $\partial X^\nu D_{\mu_1}
F_{\nu}^{\quad\mu_3} F_{\mu_3}^{\quad\mu_4} F_{\mu_5}^{\quad\mu_1}
F_{\mu_4}^{\quad\mu_5} $. }
  \label{fig_open_diagram_2}
\end{figure}

Firstly we compute diagram of figure.~\ref{fig_open_diagram_2}, which has the vertex structure$\partial X^\nu D_{\mu_1}
F_{\nu}^{\quad\mu_3} F_{\mu_3}^{\quad\mu_4} F_{\mu_5}^{\quad\mu_1}
F_{\mu_4}^{\quad\mu_5} $.
 The
 open string integral is  
\begin{equation}\label{eq:42Cplus}
I_{42, C_+} = \int_{-\infty}^{\infty} d t_4 \int_{-\infty<t_1<t_2<t_3<t_4}dt_1 dt_2 dt_3 \frac{\ln{t_{21}}}{t_{31}
    t_{42}t_{43}}.
\end{equation}
Using the change of variables eq.~\eqref{eq:41CplusCOV}, we get
\begin{align}
  \label{eq:42CplusRes}
I_{42, C_+}   &= \int_{-\infty}^{\infty} dt_4 \int_{-\infty}^0 dt_{14} \int_{t_{14}}^0 dt_{24} \int_{t_{24}}^0 dt_{34}
      \frac{\ln{t_{41} - t_{42}}}{(t_{41}-t_{43})
    t_{42} t_{43}} \nonumber \\
    &= \int_{-\infty}^{\infty} dt_4   \int_0^{\infty} \frac{dw}{w} \int_{0}^{1} dv \int_0^1 du
       \frac{\ln{w}  + \ln{(1-v)}}{u v (1-u v)} \nonumber\\
     &=  \int_{-\infty}^{\infty} dt_4 ( -2 \zeta_3 (\ln{\lambda} - \ln{\epsilon})).
\end{align}

The  heterotic string integral needs the other contour  
\begin{equation}\label{eq:42Cminus}
I_{42, C_-} = \int_{\infty}^{-\infty} d t_1 \int_{\infty>t_4>t_3>t_2>t_1}dt_4 dt_3 dt_2 \frac{\ln{t_{21}}}{t_{31}
    t_{42}t_{43}}.
\end{equation}
Using the change of variables eq.~\eqref{eq:41CminusCOV}
we get
\begin{align}
  \label{eq:42CminusRes}
I_{42, C_-}   &= - \int_{-\infty}^{\infty} dt_1 \int_{\infty}^0 dt_{41} \int_{t_{41}}^0 dt_{31} \int_{t_{31}}^0 dt_{21}
       \frac{\ln{t_{21}}}{t_{31}
    (t_{41} - t_{21}) (t_{41}- t_{31})} \nonumber \\
    &= \int_{-\infty}^{\infty} dt_1   \int_0^{\infty} \frac{dw}{w} \int_{0}^{1} dv \int_0^1 du
       \frac{\ln{w}  + \ln{v} + \ln{u}}{ (1-u v) (1-v)} \nonumber\\
     &=  \int_{-\infty}^{\infty} dt_1 ( -\zeta_3 (\ln{\lambda} - \ln{\epsilon})).
\end{align}
The heterotic integral is the sum of eq.~\eqref{eq:42CplusRes} and eq.~\eqref{eq:42CminusRes}, while the open string integral is just eq.~\eqref{eq:42CplusRes}. We see that the sv-map for $\zeta_3$ is satisfied $sv(\zeta_3) = (\frac{3}{4}) \times 2 \zeta_3$. 

\subsubsection{Case 3}
\label{sec:case_3}
\begin{figure}
  \centering
  \includegraphics[width=10cm]{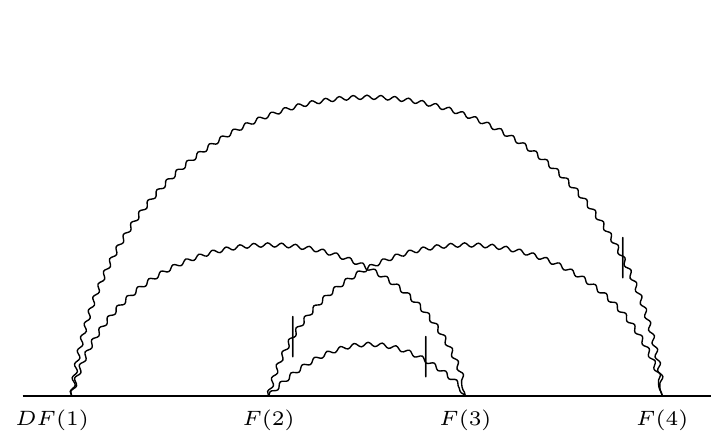}
  \caption{The Feynman diagram corresponding to structure $\partial X^\nu
D_{\mu_1} F_{\nu}^{\quad\mu_3}F_{\mu_4}^{\quad\mu_5} F_{\mu_3}^{\quad\mu_4}
F_{\mu_5}^{\quad\mu_1}$.  }
  \label{fig_open_diagram_3}
\end{figure}

Firstly we compute diagram of figure.~\ref{fig_open_diagram_3}, which has the vertex structure $\partial X^\nu
D_{\mu_1} F_{\nu}^{\quad\mu_3}F_{\mu_4}^{\quad\mu_5} F_{\mu_3}^{\quad\mu_4}
F_{\mu_5}^{\quad\mu_1}$. 
 The
 open string integral is  
\begin{equation}\label{eq:43Cplus}
I_{43, C_+} = \int_{-\infty}^{\infty} d t_4 \int_{-\infty<t_1<t_2<t_3<t_4}dt_1 dt_2 dt_3 \frac{\ln{t_{31}}}{t_{41}
    t_{42}t_{32}}.
\end{equation}
Using the change of variable eq.~\eqref{eq:41CplusCOV}
we get
\begin{align}
  \label{eq:43CplusRes}
I_{43, C_+}   &= \int_{-\infty}^{\infty} dt_4 \int_{-\infty}^0 dt_{14} \int_{t_{14}}^0 dt_{24} \int_{t_{24}}^0 dt_{34}
      \frac{\ln{t_{41} - t_{43}}}{t_{41}
    t_{42} (t_{42} - t_{43})} \nonumber \\
    &= \int_{-\infty}^{\infty} dt_4   \int_0^{\infty} \frac{dw}{w} \int_{0}^{1} dv \int_0^1 du
       \frac{\ln{w}  + \ln{(1-u v)}}{v (1-u)} \nonumber\\
     &=  \int_{-\infty}^{\infty} dt_4 ( 2 \zeta_3 (\ln{\lambda} - \ln{\epsilon})).
\end{align}

The
 heterotic string integral needs the other contour  
\begin{equation}\label{eq:43Cminus}
I_{43, C_-} = \int_{\infty}^{-\infty} d t_1 \int_{\infty>t_4>t_3>t_2>t_1}dt_4 dt_3 dt_2 \frac{\ln{t_{31}}}{t_{41}
    t_{42}t_{32}}.
\end{equation}
Using the change of variables eq.~\eqref{eq:41CminusCOV}
we get
\begin{align}
  \label{eq:43CminusRes}
I_{43, C_-}   &= - \int_{-\infty}^{\infty} dt_1 \int_{\infty}^0 dt_{41} \int_{t_{41}}^0 dt_{31} \int_{t_{31}}^0 dt_{21}
      \frac{\ln{t_{31}}}{t_{41}
   (t_{41}-t_{21}) (t_{31} - t_{21})}\nonumber \\
    &= \int_{-\infty}^{\infty} dt_1   \int_0^{\infty} \frac{dw}{w} \int_{0}^{1} dv \int_0^1 du
       \frac{\ln{w}  + \ln{v}}{(1-u) (1-u v)} \nonumber\\
     &=  \int_{-\infty}^{\infty} dt_1 ( \zeta_3 (\ln{\lambda} - \ln{\epsilon})).
\end{align}
The heterotic integral is the sum of eq.~\eqref{eq:43CplusRes} and eq.~\eqref{eq:43CminusRes}, while the open string integral is just eq.~\eqref{eq:43CplusRes}. We see that the sv-map for $\zeta_3$ is satisfied $sv(\zeta_3) =(\frac{3}{4}) \times  2 \zeta_3$. 

So from our computation, we have $sv(\zeta_3) =(\frac{3}{4}) \times  2 \zeta_3$.  There is a total factor of $3/4$ here, compared with the mathematical result. We argue that this factor is just a total factor for all the diagrams at four loop level, so it can be incorporated into the action, since we found this same factor in both section~\ref{sec:case_2} and section~\ref{sec:case_3}.

\section{Conclusion}

In this paper, we address the sv-map from the nonlinear sigma model approach. We show that the sv-map comes from a sum of two opposite-directed integral contours, when the gauge physics of both the open and the heterotic string sigma models are under the Wilson loop representation. In referene ~\cite{FFST2018195}, the sv-map is shown to come from a sum of six radial orderings of heterotic vertices on the complex plane, when the gauge physics of the heterotic sigma model is not under the Wilson loop representation. So the Wilson loop representation gives sv-map a simpler geometric origin. 
To do that, we build a Wilson loop for the heterotic string sigma model and prove that it is the exact propagator of the fermion field that carry the gauge physics of the heterotic string in the fermionic represenation. Then we construct the action of the heterotic sigma model using this Wilson loop, by studying the goemetry of Wilson loop and generalizing the nonabelian Stokes's theorem into the fermionic case.  We have shown how the sv-map arises for $\zeta_2$ and $\zeta_3$ at three loop and four loop level, from the sum of contours of the Wilson loop representation. Based on these, we finally conjecture that the sv-map of a general MZV  comes from this sum of contours of the Wilson loop representation. 

\section{Acknowledgments}
The author is grateful to Stephan Stieberger for helpful suggestions and conversations. The author is also grateful to 
 Angelos Fotopoulos, Zygmunt Lalak,  and Tomasz R. Taylor for communications. This material is based in part upon work supported by the National Science Foundation under Grant Number PHY—1620575. Any opinions, findings, and conclusions or recommendations	expressed in this material are those of the author and do not necessarily reflect the views of the National Science Foundation.

\section*{References}
% Create the reference section using BibTeX:
\bibliography{wilson_loop_construction}

\end{document}